\documentclass[12pt]{article}
\usepackage[dvips]{graphicx}
\usepackage{epsfig,cite}
\usepackage {amsmath}
\usepackage{color}
\usepackage{amssymb}
\usepackage{slashed}
\include{epsf}
\newcount\timecount
\newcount\hours \newcount\minutes  \newcount\temp \newcount\pmhours
\hours = \time
\divide\hours by 60
\temp = \hours
\multiply\temp by 60
\minutes = \time
\advance\minutes by -\temp
\def\hour{\the\hours}
\def\minute{\ifnum\minutes<10 0\the\minutes
            \else\the\minutes\fi}
\def\clock{
\ifnum\hours=0 12:\minute\ AM
\else\ifnum\hours<12 \hour:\minute\ AM
      \else\ifnum\hours=12 12:\minute\ PM
            \else\ifnum\hours>12
                 \pmhours=\hours
                 \advance\pmhours by -12
                 \the\pmhours:\minute\ PM
                 \fi
            \fi
      \fi
\fi
}

\def\monthname{\relax\ifcase\month 0/\or January\or February\or
   March\or April\or May\or June\or July\or August\or September\or
   October\or November\or December\else\number\month/\fi}

\def\bold#1{\setbox0=\hbox{$#1$}%
     \kern-.025em\copy0\kern-\wd0
     \kern.05em\copy0\kern-\wd0
     \kern-.025em\raise.0433em\box0 }

\def\beq{\begin{equation}}
\def\eeq{\end{equation}}

\def\ga{\mathrel{\raise.3ex\hbox{$>$\kern-.75em\lower1ex\hbox{$\sim$}}}}
\def\la{\mathrel{\raise.3ex\hbox{$<$\kern-.75em\lower1ex\hbox{$\sim$}}}}
\def\gev{{\rm \, Ge\kern-0.125em V}}
\def\tev{{\rm \, Te\kern-0.125em V}}
\def\gyr{{\rm \, G\kern-0.125em yr}}

\def\gappeq{\mathrel{\rlap {\raise.5ex\hbox{$>$}}
{\lower.5ex\hbox{$\sim$}}}}
\def\lappeq{\mathrel{\rlap{\raise.5ex\hbox{$<$}}
{\lower.5ex\hbox{$\sim$}}}}
\def\Toprel#1\over#2{\mathrel{\mathop{#2}\limits^{#1}}}

\def\m12{m_{1\!/2}}

\def\bea{\begin{eqnarray}}
\def\eea{\end{eqnarray}}

\def\beqar{\begin{eqnarray}}
\def\eeqar{\end{eqnarray}}

\begin{document}

\begin{titlepage}
\pagestyle{empty}
\baselineskip=21pt
\rightline{}
\vskip 0.8in
\begin{center}
{\Large\bf{Exploring the Anomalous Higgs-top Couplings}}
\end{center}
\begin{center}
\vskip 0.4in
{\bf Sara Khatibi and Mojtaba Mohammadi Najafabadi }
\vskip 0.1in
{\it  School of Particles and Accelerators, \\
Institute for Research in Fundamental Sciences (IPM) \\
P.O. Box 19395-5531, Tehran, Iran}\\
\vspace{2cm}
 \textbf{Abstract}\\
 \end{center}
\baselineskip=18pt \noindent
{
Top quark with its large Yukawa coupling is crucially important to explore TeV scale
physics. Therefore, the study of
Higgs-top sector is highly motivated to look
for any deviations from the standard model predictions.
The most general lowest order Lagrangian for the Higgs-top Yukawa coupling has scalar 
($\kappa$) and pseudoscalar ($\tilde{\kappa}$) components. Currently, these couplings
are constrained indirectly using the present experimental limits on the Higgs-$\gamma$-$\gamma$
and Higgs-gluon-gluon couplings. Furthermore, stronger bounds on $\kappa$ and $\tilde{\kappa}$
are obtained using the limits on the electric dipole moments (EDM). In this work, we 
propose an asymmetry-like observable $O_{\phi}$ in $t\bar{t}H$ production at the LHC
to probe the Higgs-top coupling and to distinguish 
between the scalar and pseudoscalar components.  We also show that the
presence of the pseudoscalar component in the Higgs-top Yukawa coupling 
leads to a sizeable value for the top quark EDM. It is shown that a limit of $10^{-19}$ e.cm,  which 
is achievable by the future $e^{-}e^{+}$ collider,  allows us to exclude a significant 
region in the $(\kappa,\tilde{\kappa})$ plane. 
}

\vfill
\leftline{PACS numbers: 14.65.Ha,14.80.Bn}
\end{titlepage}
\baselineskip=18pt


\section{Introduction}

In the framework of the Standard Model (SM) the Higgs boson gives masses to 
fermions (charged leptons and quarks) through Yukawa couplings. The couplings of the
Higgs boson to the gauge bosons and fermions have measured in different decay modes. 
The Higgs  couplings to $W$ and $Z$ bosons are measured with the uncertainties of
around $20\%-30\%$ and the Higgs couplings to the top and bottom quarks
are known with the uncertainties of $30\%$ and $40\%$, respectively \cite{errors}.
The couplings are in agreement 
with the SM predictions within these relatively large uncertainties. Top quark which is the 
heaviest particle among the SM particles has the largest coupling with the Higgs boson.
The precise measurement of the Higgs-top Yukawa interaction enables us to check 
the spontaneous symmetry breaking mechanism closely and in case of observing any deviation
it is a window to new physics beyond the SM.
So far, there are many attempts in the way of probing top Yukawa couplings at the colliders using 
$t\bar{t}H$ production, single top plus Higgs boson production and via the Higgs pair production \cite{th1},\cite{th2},
\cite{th3},\cite{th4},\cite{th5},\cite{th6},\cite{th7},\cite{th8}.
There are also many studies to explore the standard and exotic Higgs couplings by using 
the available data and by proposing new sensitive observables to achieve precise measurements 
and to look for deviations from the SM
\cite{h1},\cite{h2},\cite{h3},\cite{h4},\cite{h5},\cite{h6},\cite{h7},\cite{h8},\cite{h9},
\cite{h10},\cite{h11}. 

Within the SM, the Higgs-top coupling is described by only purely scalar type of coupling.
However, in models beyond the SM the Higgs-top coupling can consists of both scalar and 
pseudoscalar components. In these models the Higgs boson can be a CP-mixed state \cite{bsm1},\cite{bsm2}.
It also should be noted that based on the present data
the admixture of the scalar and pseudoscalar couplings is still possible \cite{th1},\cite{th7}.
The generic form of the Higgs-top Yukawa coupling can be parametrized as \cite{lag}:
\begin{eqnarray}\label{lag}
\mathcal{L}=-\frac{m_{t}}{v}\bar{t}(\kappa+i \tilde{\kappa} \gamma_{5})t H + h.c.
\label{lag}
\end{eqnarray}
where $m_{t}$ is the top quark mass and $v$ denotes the vacuum expectation value. The parameters
$\kappa$ and $\tilde{\kappa}$ are dimensionless real parameters. In the SM at the leading order (LO),
the value of $\kappa=1$ and $\tilde{\kappa}=0$. It is notable that the CP-violating component, $\tilde{\kappa}$,
can arise from loops at higher orders in the SM which is expected to be small.  
At present, there are indirect limits on $\kappa$ and $\tilde{\kappa}$ from the experimental 
measurements of $H-\gamma-\gamma$ and Higgs-gluon-gluon couplings and from the upper bounds on 
the electron, neutron, etc. electric dipole moments (EDM) \cite{limits1},\cite{limits2}.
Among the electron, neutron and mercury electric dipole moments, the upper limit
on the electron EDM gives the most stringent limit on $\tilde{\kappa}$. The contribution 
of the CP-violating component of the top-Higgs Yukawa coupling to the electron EDM 
is given by:
\begin{eqnarray}
d_{e} = 9\times 10^{-27} \tilde{\kappa}~e.cm
\end{eqnarray}
where the electron Yukawa coupling is taken to be equal to the SM value. This
leads to an upper bound of 0.01 on $\tilde{\kappa}$. The production of the 
Higgs boson through gluon fusion and the Higgs decay into di-photon 
are affected by the CP-vilating component of the Higgs-top coupling $\tilde{\kappa}$.
The ratio of the cross section of the Higgs production in gluon fusion 
considering CP-violating component to the SM cross section has been found to be:
\begin{eqnarray}
\mu_{gg}(\kappa,\tilde{\kappa}) \simeq 1.11\kappa^{2}+ 2.6\tilde{\kappa}^{2} -0.11\kappa 
\end{eqnarray}
As it can be seen, the cross section of Higgs production via gluon fusion
is more affected by the CP-violating term than the SM component. Also, 
the CP-violating term has positive contribution and always lead to enhance 
the Higgs production rate (or to the decay of Higgs to gluon-gluon). 
In the Higgs boson decay into di-photon
the ratio of the $H\rightarrow \gamma \gamma$ in the presence of
$\tilde{\kappa}$ to the SM width has the following form:
\begin{eqnarray}
\mu_{\gamma \gamma}(\kappa,\tilde{\kappa}) \simeq 0.078\kappa^{2} -0.71\kappa +  0.18\tilde{\kappa}^{2} +1.6
\end{eqnarray}
Similar to the Higgs production in gluon fusion the CP-violating coupling
has a constructive effect on the rate of Higgs decay into di-photon and
leads to an enhancement. However, the power of enhancement is weaker than
the Higgs-gluon-gluon coupling. A global analysis including 
several observables provides the following best-fit values for the anomalous
parameters $\kappa$ and $\tilde{\kappa}$ \cite{th7}:
\begin{eqnarray}
\kappa,\tilde{\kappa} \longrightarrow 0.8,\pm 0.3 \nonumber
\end{eqnarray}
This means that non-zero values for the CP-violating coupling $\tilde{\kappa}$
are allowed. It is remarkable that $\tilde{\kappa}$ is allowed to 
vary in the range of -0.4 to 0.4 when $\kappa$ is fixed to unity \cite{th7}.
In this paper, the main idea is to develop a sensitive kinematic 
observable to the Higgs-top Yukawa coupling.

As mentioned previously, there are several studies at colliders to examine the Higgs-top Yukawa couplings.
For example, in \cite{th111} the authors have proposed an asymmetry in the charged
lepton energy in top pair events at hadron colliders. It has been shown that such an
asymmetry is sensitive to the CP violating term $\tilde{\kappa}$ and is large enough
above the backgrounds.  In \cite{th1}, the sensitivity of the  total
cross sections of the $t\bar{t}H$, $tH$, and $\bar{t}H$ processes have
been examined to the Higgs-top Yukawa couplings. It has been found
that the total cross section of $t\bar{t}H$ process decreases significantly with
increasing the ratio of $\tilde{\kappa}/\kappa$. In addition, it has
been shown that the distribution of $\Delta\phi_{l^{+}l^{-}}$ (the
difference between the azimuthal angle of the charged leptons in top
quarks decay) is sensitive to both the sign and the magnitude of $\tilde{\kappa}/\kappa$.

In the next section, based on the angular distributions of the final state particles in 
$pp\rightarrow t\bar{t}H$ process an asymmetry-like observable $O_{\phi}$ is defined. We show that
this observable considerably sensitive to the CP violating term of the Higgs-top coupling. 
We examine the effect of the main background and kinematic cuts on this observable. Then, 
the sensitivity of $O_{\phi}$ at the LHC with the center-of-mass energy of 14 TeV as a function 
of the integrated luminosity is presented. In section 3, we show that the general 
Higgs-top couplings generates a sizeable EDM for the top quark. Then, using a possible 
future bound of $10^{-19}$ e.cm  \cite{topedm} on the top EDM,  which is achievable by the future electron-positron 
collider,  an allowed region in $\kappa,\tilde{\kappa}$ is obtained. Finally, section 4
concludes the paper.

\section{Azimuthal angular observable}

We construct an asymmetry-like observable from the azimuthal 
angular distributions of the final state objects in $pp\rightarrow t\bar{t}H$ process. This 
observable, $O_{\phi}$, is a sensitive discriminant to distinguish between the
CP-conserving ($\kappa$) and CP-violating ($\tilde{\kappa}$) parts of the $t\bar{t}H$ 
couplings.

We define an angular asymmetry-like $O_{\phi}$ with respect to the azimuthal angle differences
$\Delta\phi_{t\bar{t}} = \phi_{t}-\phi_{\bar{t}}$ and $\Delta\phi_{tH} = \phi_{t}-\phi_{H}$ as:
\begin{eqnarray}
O_{\phi}=\frac{N(|\Delta \phi(t \bar{t})|>|\Delta \phi(t H)|)-N(|\Delta \phi(t \bar{t})|<|\Delta \phi(t H)|)}{N(|\Delta \phi(t \bar{t})|>|\Delta \phi(t H)|)+N(|\Delta \phi(t \bar{t})|<|\Delta \phi(t H)|)}
\end{eqnarray}
Here the first (second) term in the nominator denotes the number 
of $t\bar{t}H$ events with $|\Delta \phi(t \bar{t})|>|\Delta \phi(t H)|$ ($|\Delta \phi(t \bar{t})|<|\Delta \phi(t H)|$). The denominator 
is the total number of events.

Now we would like to examine  our proposed observable $O_{\phi}$.
To generate the hard scattering matrix elements the {\sc MadGraph} 5 package \cite{Alwall:2011uj}
is used  with the CTEQ6L \cite{Pumplin:2002vw} as the proton
parton distribution function (PDF).
In our analysis, we concentrate on the LHC run with the center-of-mass energy of $\sqrt{s}=14$ TeV.
In order to simulate the signal events, i.e. $pp\rightarrow t\bar{t}H$ with the general 
form of the $t\bar{t}H$, the effective Lagrangian of Eq.\ref{lag}
is implemented within the {\sc  FeynRules} package \cite{Christensen:2008py},\cite{Duhr:2011se}. Then 
the model  is imported  to a UFO module \cite{Degrande:2011ua} and then inserted to the {\sc MadGraph} 5.

\begin{figure}
\centering
\includegraphics[width=7cm,height=6cm]{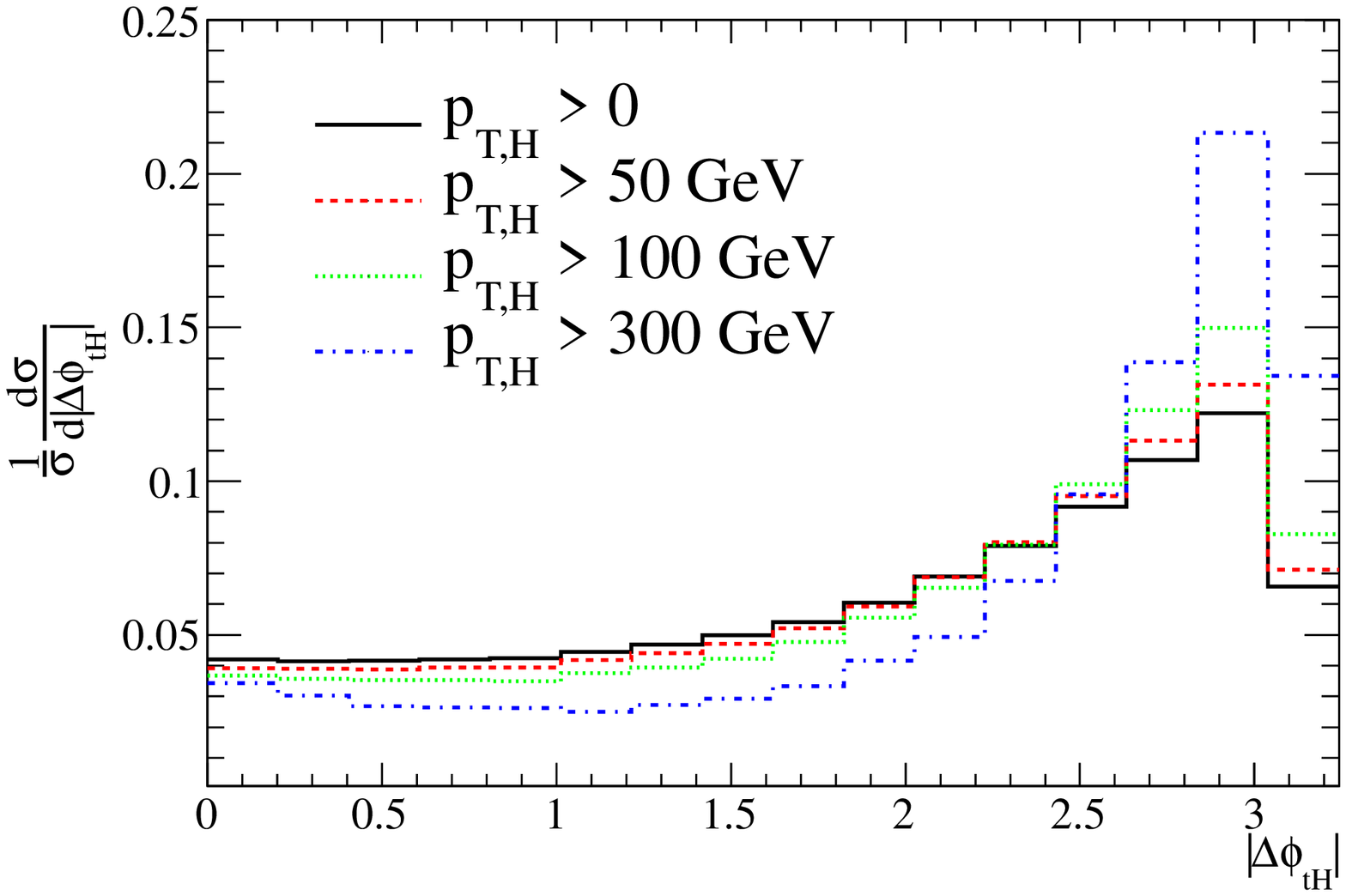}
\includegraphics[width=7cm,height=6cm]{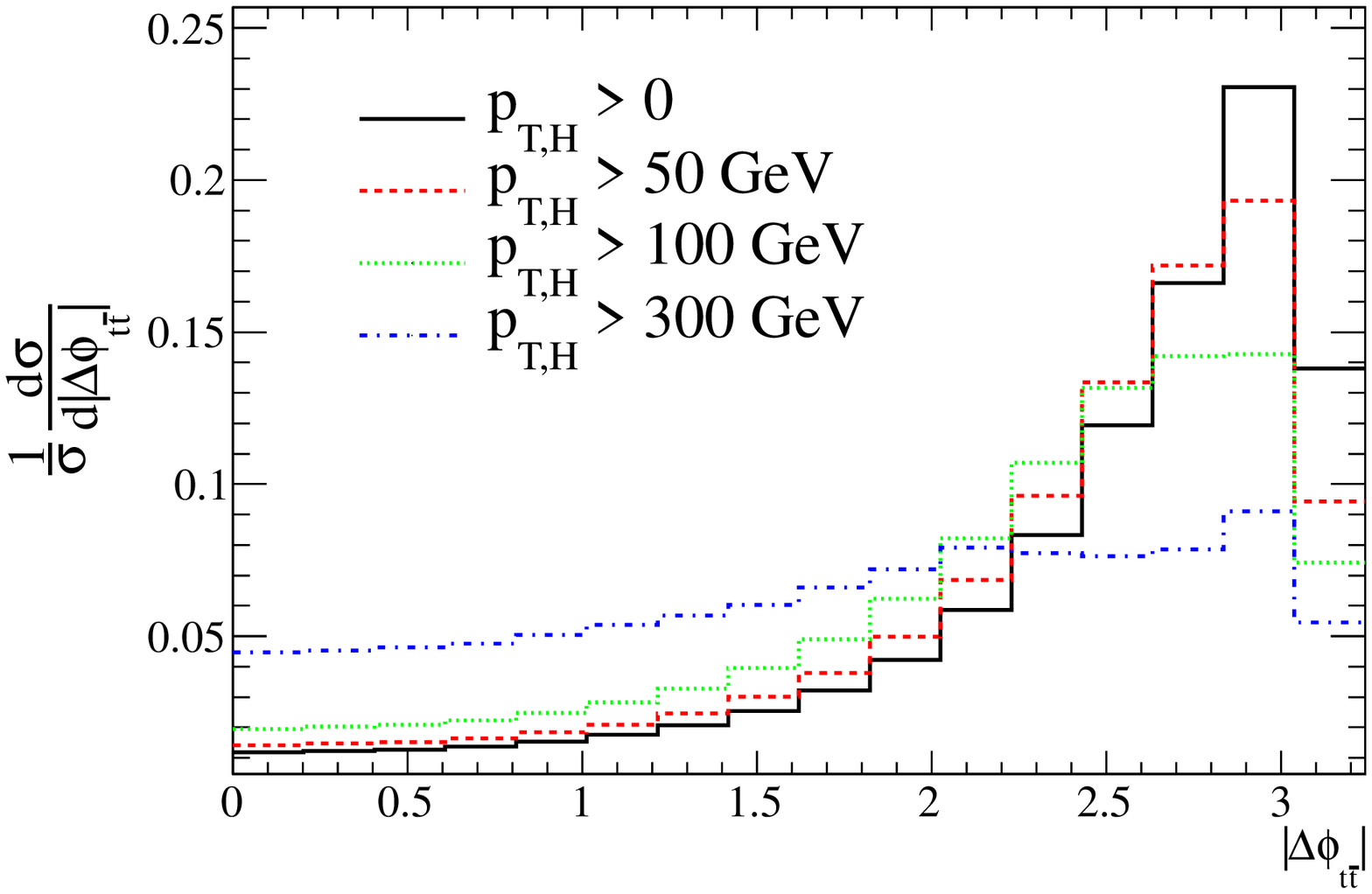}
\caption{The distributions of the angular separations of $t\bar{t}$ and $tH$ for $t\bar{t}H$
final state in proton-proton collisions at $\sqrt{s}=14$ TeV for different values of cut on the transverse Higgs momentum. }
\label{dphi}
\end{figure}

We show the values of $O_{\phi}$ for the SM case ($\kappa = 1$ and $\tilde{\kappa}=0$)
for different cuts on the Higgs boson transverse momentum in Table \ref{oSM}.
As it can be seen the value of $O_{\phi}$ decreases when the cut on the Higgs-$p_{T}$ grows.
To understand the reason for such a behavior, in Fig.\ref{dphi} we plot parton level 
distributions of $\Delta\phi_{t\bar{t}}$ and $\Delta\phi_{tH}$ for example for the
SM case ($\kappa=1,\tilde{\kappa}=0$) with different cuts on the Higgs boson transverse momentum.
When no cut is applied on the Higgs-$p_{T}$, the top and anti-top quarks prefer to fly 
in opposite directions in the transverse plane. However, with increasing the Higgs 
$p_{T}$ cut the $\Delta\phi_{t\bar{t}}$ distribution tends to become flat and instead the top quark
and the Higgs boson move to a back-to-back position in the transverse plane.
As shown in Fig.\ref{dphi}, the number of the events with $|\Delta \phi(t \bar{t})|>|\Delta \phi(t H)|$ decreases 
when we impose a larger cut on the Higgs boson transverse momentum.

There are  sources of theoretical uncertainties to the observable $O_{\phi}$.
The uncertainty originating from the variation of the factorization and renormalization scales 
 is obtained by varying the two scales together in the range of $\mu=\mu_{0}/2$  to 
 $\mu=2\mu_{0}$ with the central scale $\mu_{0}$ is set to
 $\sqrt{p_{T}^{2} + m_{t}^{2}}$.
  The uncertainty coming from the limited knowledge of the parton distribution functions PDF
 is  calculated using  the 44
members of the CTEQ6.6 PDFs  \cite{pdf}.
 There is an uncertainty from the top quark mass which is obtained 
 by variation of the top quark mass $\pm 1$ GeV.
 The relative uncertainties from variation of the factorization/renormalization scales, PDF, 
 and top quark mass are found to be $1\%$,   $0.3\%$, and $0.09\%$, respectively.
 The dominant source of the theoretical uncertainty is from the 
 variation of the factorization/renormalization scales. These
 uncertainties are small with respect to the relative deviation from
 the CP violating term, $\tilde{\kappa}$. From Table \ref{oSM}, the
 relative deviation due to the presence of CP violating term
 with $\kappa = 1, \tilde{\kappa}=0.4$ leads to around $10.95\%$
change with respect to the SM value.

\begin{table} 
\begin{center}
\begin{tabular}{c||c|c|c|c|c}
\hline
    cut on Higgs-$p_{T}$   (GeV)              &  0.0     &   50.0    &    100.0   &      200.0  & 300.0
 \\ \hline 
 $O_{\phi} (\kappa=1,\tilde{\kappa}=0)$    & 0.356    & 0.229    & 0.021      &    -0.265    & -0.411\\ \hline 
 $O_{\phi} (\kappa=1,\tilde{\kappa}=0.4)$    & 0.317    & 0.191    & -0.006      &    -0.299    & -0.441 \\ \hline 
\end{tabular}
\caption{ The values of the asymmetry-like observable $O_{\phi}$ for the SM and the case of $\tilde{\kappa} = 0.4$
for several cuts on the Higgs boson transverse momentum. }
\label{oSM}
\end{center}
\end{table}

\begin{figure}
\centering
  \includegraphics[width=7cm,height=5cm]{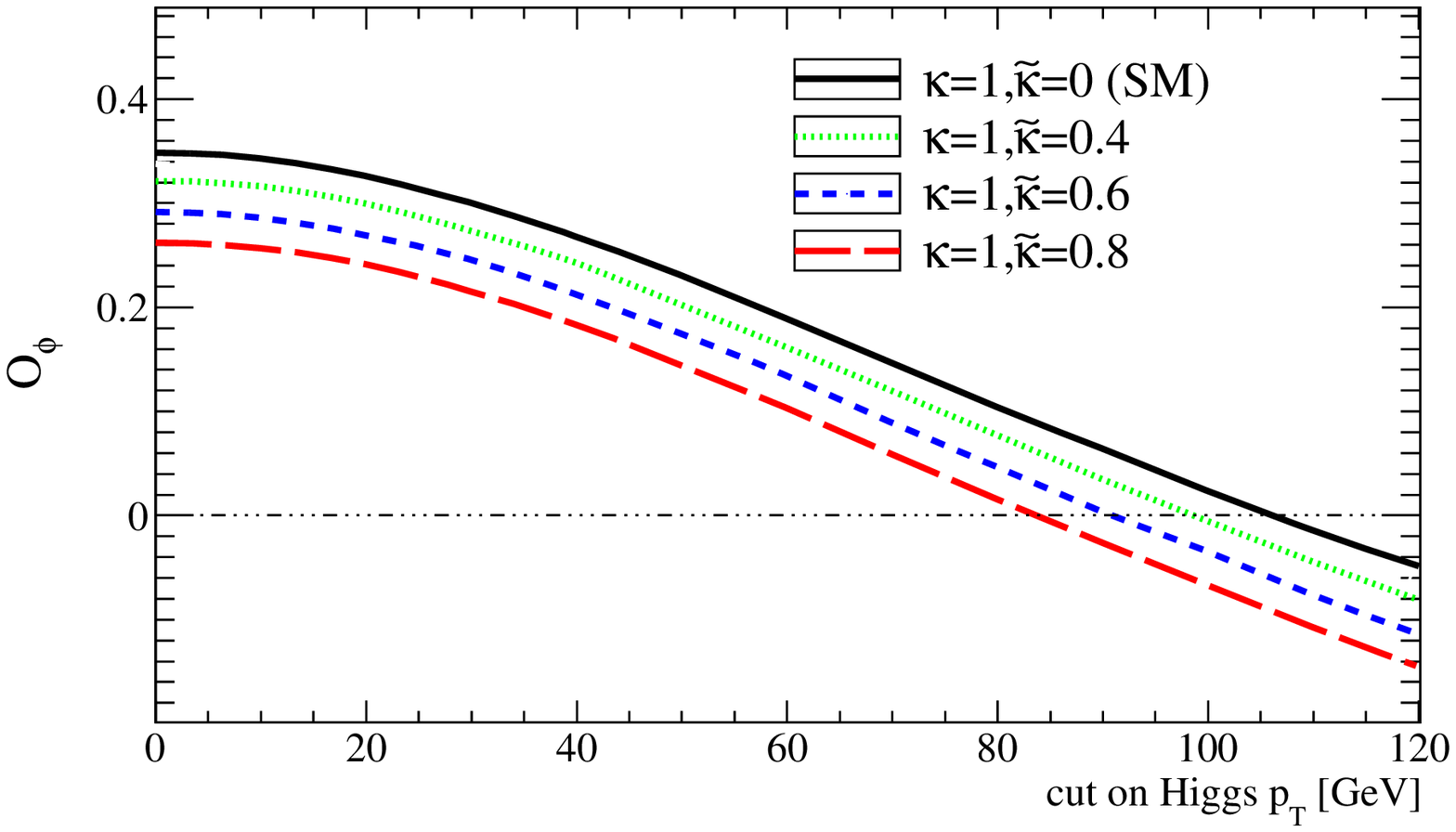}
  \includegraphics[width=7cm,height=5cm]{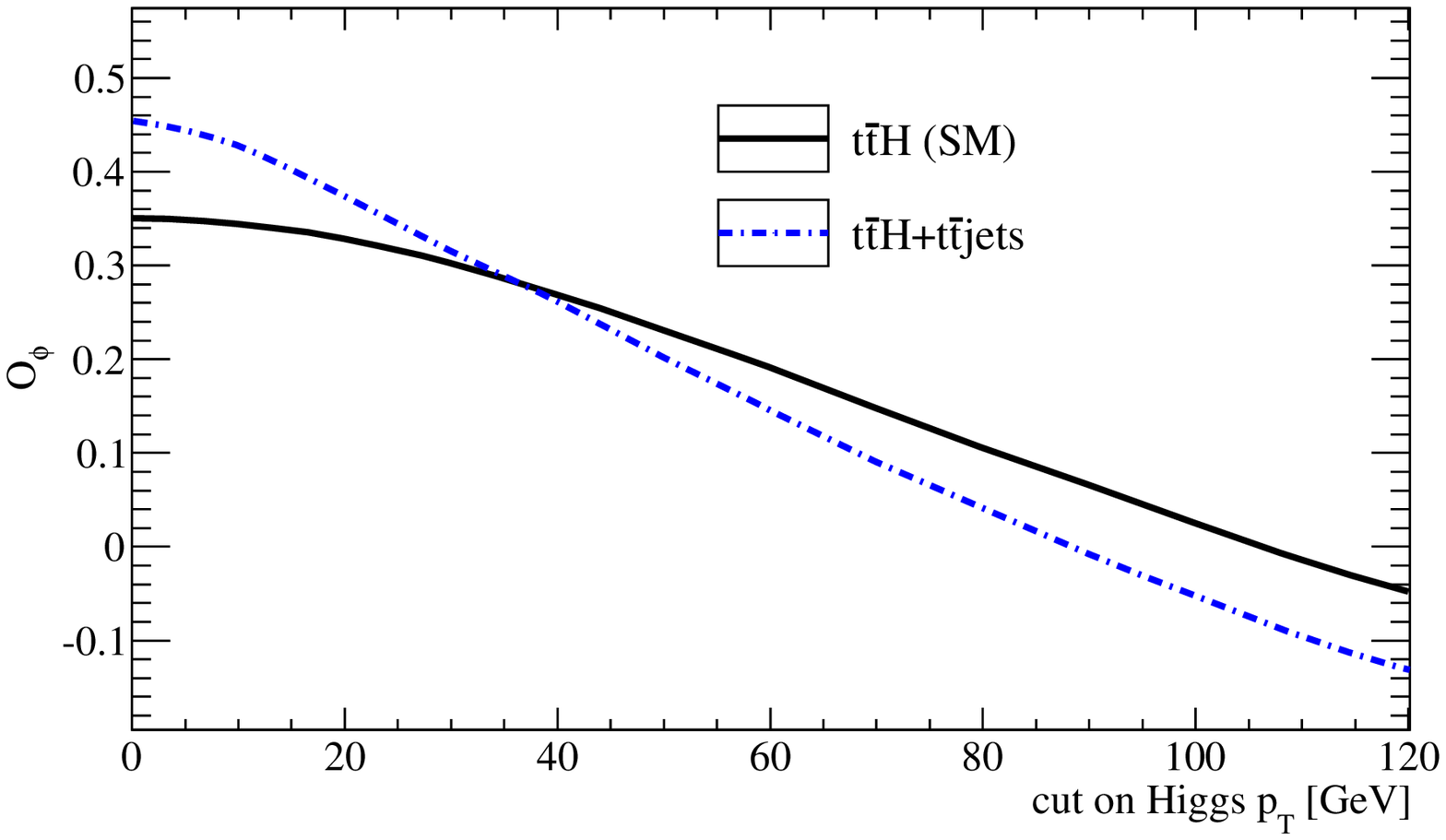}
  \caption{Left: the shape of $O_{\phi}$ for the SM and several values of $\tilde{\kappa}$
versus cuts on the transverse Higgs momentum. Right: the shape of $O_{\phi}$ of $t\bar{t}H$ for the SM
case and for the $t\bar{t}H$ plus the main background $t\bar{t}$+jets.}\label{o_phi}
\end{figure}

Now, we switch to our general effective Lagrangian, Eq.\ref{lag}, with arbitrary
values of $\kappa$ and $\tilde{\kappa}$. In the left panel of Fig.\ref{o_phi}, we depict the shape
of $O_{\phi}$ observable for $\kappa=1$ and various values of $\tilde{\kappa}$
as a function of the Higgs boson $p_{T}$. The presence of a CP-violating 
coupling in $t\bar{t}H$ interaction reduces the amount of  $O_{\phi}$ at any value of the cut on Higgs boson $p_{T}$. 

Notably the variation of $\kappa$ in the range of for example 0.0 to 1.0 at a given value 
of $\tilde{\kappa}$ leads to no significant deviation in $O_{\phi}$. The deviation 
is of the order of $1\%$. This indicates that our azimuthal angular based
observable is only sensitive to the CP-violating part of the $t\bar{t}H$ coupling and can be 
used to probe $\tilde{\kappa}$ coupling. Therefore, such an observable allows
us to distinguish between scalar and pseudoscalar couplings in $t-\bar{t}-H$ interaction.

The amount of observable $O_{\phi}$ drops for the case that the Higgs bosons have
larger transverse momentum. It tends to zero at the cut value of around 105 GeV
on the Higgs boson $p_{T}$ for the SM $(\kappa=1,\tilde{\kappa}=0)$. The Higgs transverse
momentum at which $O_{\phi}=0$ varies for different values of $\tilde{\kappa}$.
The larger value of $\tilde{\kappa}$ is corresponding to lower Higgs $p_{T}$
at which $O_{\phi}$ becomes zero. This phenomena happens when there is 
a balance between the number of events with $|\Delta\phi_{t\bar{t}}|>|\Delta\phi_{tH}|$
and $|\Delta\phi_{t\bar{t}}|<|\Delta\phi_{tH}|$. The value of cut on Higgs transverse momentum
that leads zero for $O_{\phi}$ could be a criteria to probe the $t\bar{t}H$ coupling. 
It is noticeable that including the backgrounds, cuts and detector effects change the 
situation.

In this study, we only concentrate on the lepton+jets decay mode 
of the $t\bar{t}$ ($t\bar{t}\rightarrow l\nu q\bar{q}'b\bar{b}$) and consider
the Higgs boson decay into a bottom-quark pair (other Higgs decays can be included).
Background estimations by the CMS experiment for the analysis of search for $t\bar{t}H$
at 7 and 8 TeV center-of-mass energies show that the main background contributions 
are originating from $t\bar{t}$+light flavor quarks, $t\bar{t}+c\bar{c}$ and $t\bar{t}+b\bar{b}$.
In the right side of Fig.\ref{o_phi}, we show that including these backgrounds change the values and shape of $O_{\phi}$
for different cuts and on Higgs $p_{T}$. 
The trend of $O_{\phi}$ versus the Higgs $p_{T}$ cut does not take significant effect 
after including the main background. Including this background even leads to larger 
slope for $O_{\phi}$.

\begin{figure}
\centering
\includegraphics[width=7cm,height=6cm]{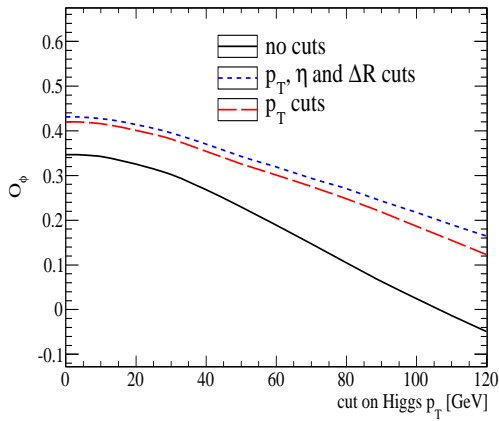}
\caption{The shape of $O_{\phi}$ before applying any kinematic cuts on the final state particles, after cuts on the
transverse momenta of lepton and jets and after applying all kinematic cuts. The trend does not change while the
amount of $O_{\phi}$ increases at any value of Higgs transverse momentum cut.}\label{cut}
\end{figure}

At this stage, we turn to impose kinematic cuts on the final state objects. 
The cuts are chosen similar to the ones used by the CMS experiment in search for 
the $t\bar{t}H$ channel \cite{cmsttH}. Charged lepton is required to have $p_{T}> 25$ GeV and 
$|\eta|<2.5$. All jets must have $p_{T}> 40$ GeV and 
$|\eta|<2.5$. In order to have well-separated objects, it is required that the 
angular separation between objects should be greater than 0.4, i.e. 
$\Delta R_{ij}=\sqrt{(\eta_{i}-\eta_{j})^{2}+(\phi_{i}-\phi_{j})^{2}}>0.4$.
Finally, the magnitude of missing transverse energy has to be larger than 30 GeV.
In Fig.\ref{cut} the effect of these cuts are presented on our observable $O_{\phi}$.
We show the shape of $O_{\phi}$ versus the cut on Higgs transverse momentum 
before applying any cut, after applying only $p_{T}$ cuts and after applying all cuts.
One can see that imposing cuts on the final state object in particular the cuts 
on transverse momenta of the final particles enhance the amount of $O_{\phi}$
with respect to the case of applying no cuts. Applying these kinematic cuts
increase the contribution of events in which the Higgs boson recoils against the
$t\bar{t}$ pair. Obviously, this effect is more visible for the events that Higgs
boson has larger boost.

The statistical significance for the observable $O_{\phi}$ is determined 
as the ratio of $O_{\phi}$ to its standard deviation $\Delta_{O_{\phi}}$:  $S=O_{\phi}/\Delta_{O_{\phi}}$.
After some algebra, one can show:
\begin{eqnarray}
\Delta_{O_{\phi}} = \sqrt{\frac{1-O_{\phi}^{2}}{\sigma_{pp\rightarrow t\bar{t}H}\times \mathcal{L}}}
\end{eqnarray}
where $\sigma_{pp\rightarrow t\bar{t}H}$ is the cross section of the $pp\rightarrow t\bar{t}H$ in the 
SM and $\mathcal{L}$ denotes the total integrated luminosity.
The signal significance $S$ in the presence of the anomalous couplings can be defined as:
\begin{eqnarray}\label{atlasbound}
S(\kappa,\tilde{\kappa})=\frac{O_{\phi}(\kappa,\tilde{\kappa})-O_{\phi}^{SM}}{\sqrt{1-(O_{\phi}^{SM})^{2}}}
\sqrt{\sigma^{SM}_{pp\rightarrow t\bar{t}H} \mathcal{L}}
\end{eqnarray}
The statistical significance $S$ is a function of $\kappa$ and $\tilde{\kappa}$. Since $O_{\phi}$
is not sensitive to $\kappa$, we set $\kappa = 1$ (its SM value) and concentrate on $\tilde{k}$.
In our analysis, we determine the value of cut on the Higgs boson $p_{T}$ that maximizes the statistical 
significance. We choose several values of $\tilde{\kappa}$ and calculate the statistical significance 
in terms of Higgs boson transverse momentum. We find that for all values of $\tilde{\kappa}$ parameter,
the maximum takes place at $p_{T,H} > 0$ i.e. no cut on the Higgs boson transverse momentum. 
This is the value at which our proposed observable takes its largest value. It is reasonable 
since when no cut is applied on the Higgs $p_{T}$ there is more statistics that leads to 
smaller statistical uncertainty and larger signal significance.

\begin{figure}
\centering
\includegraphics[width=7cm,height=6cm]{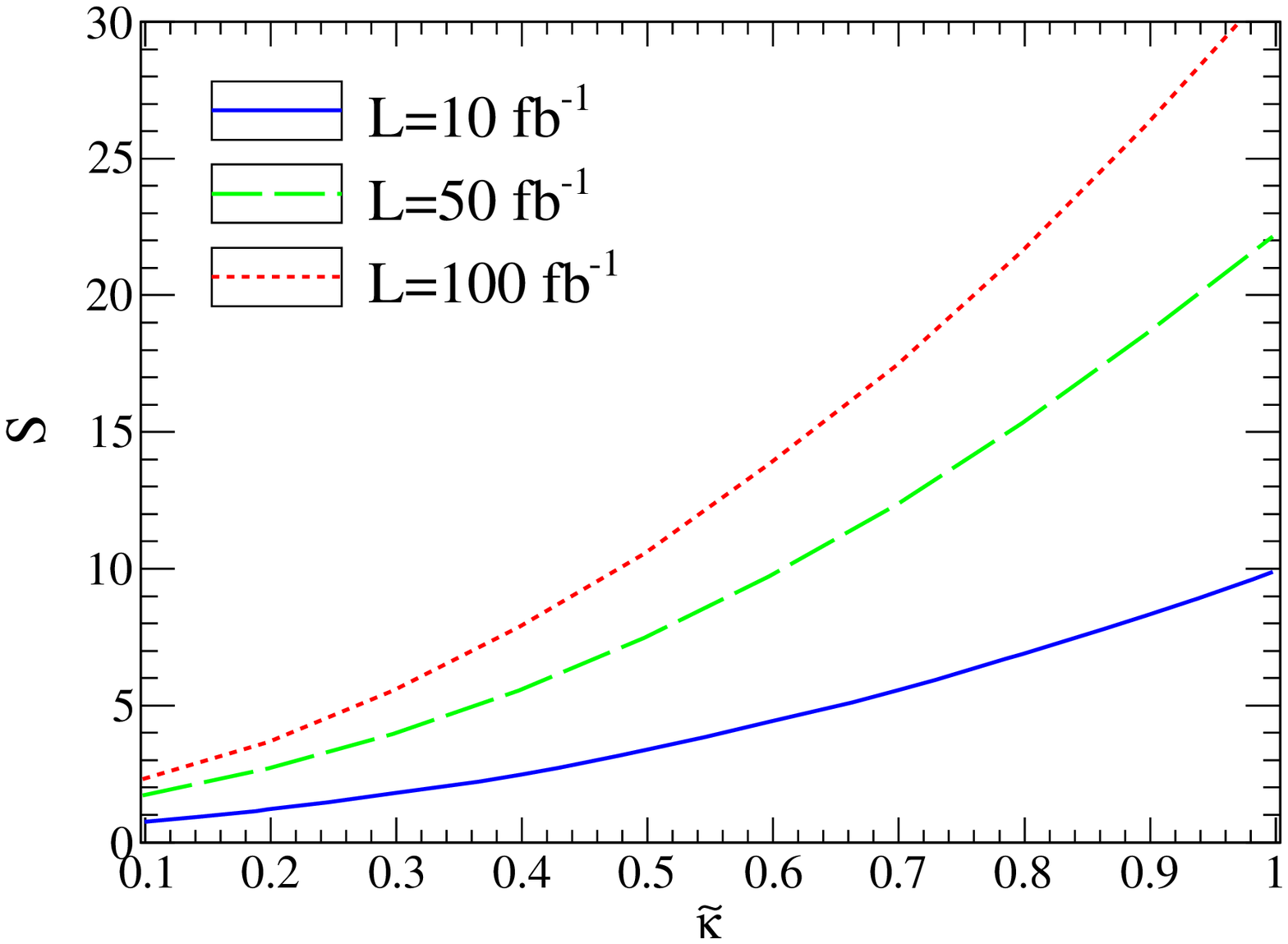}
\includegraphics[width=7cm,height=6cm]{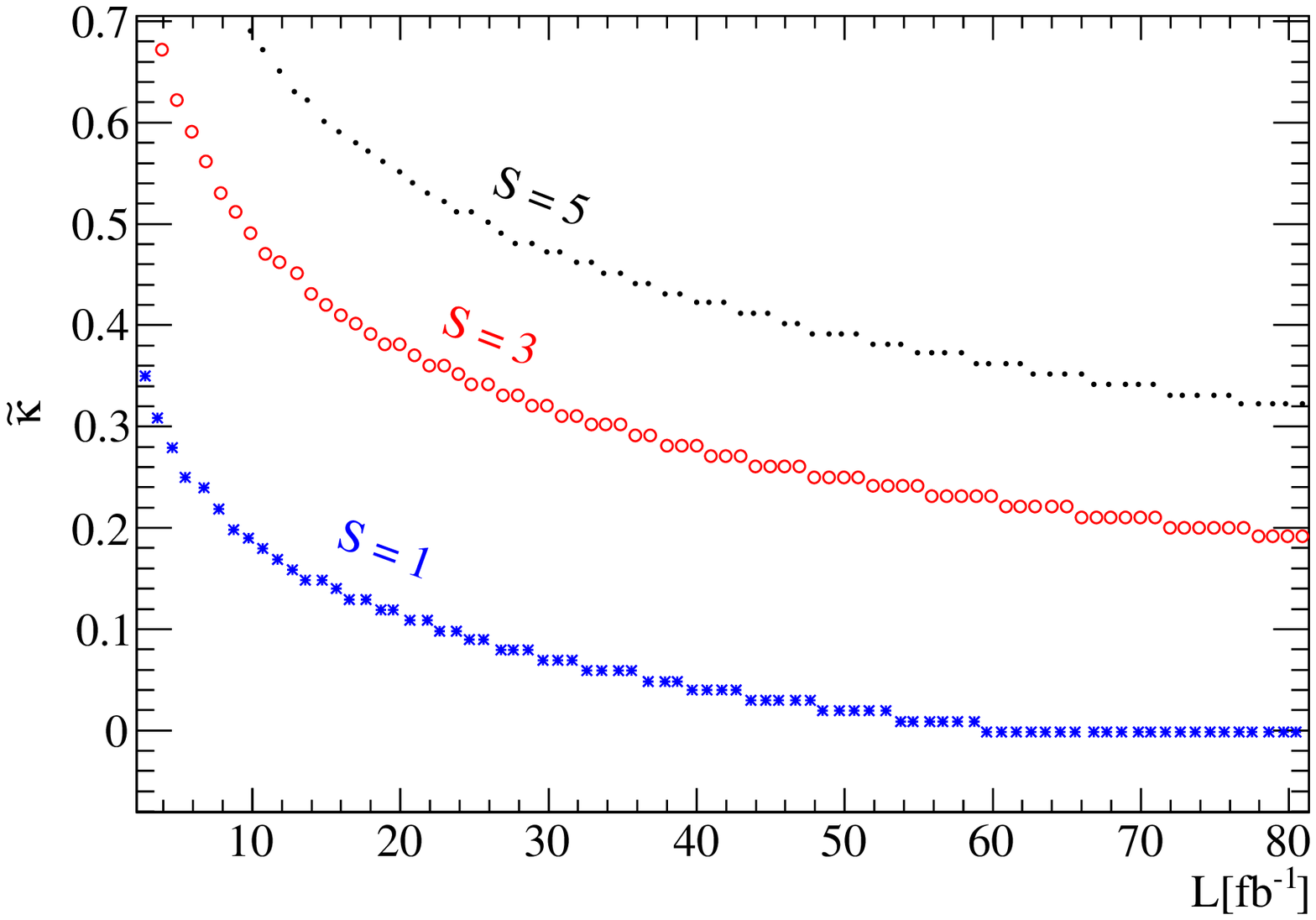}
\caption{Left: The significance as a function of $\tilde{\kappa}$ for integrated 
luminosities of 10,50,100 fb$^{-1}$.
Right:The 1,3, and 5 sigma regions of $\tilde{\kappa}$ as a function of the integrated luminosity in $pp$ collisions at the
center-of-mass energy of 14 TeV.}\label{sig}
\end{figure}

The left plot in Fig.\ref{sig} shows the statistical 
significance versus $\tilde{\kappa}$ for the integrated luminosities of 10,50,100 fb$^{-1}$ at the
LHC for the center-of-mass energy of 14 TeV. In the right side of Fig.\ref{sig}, the $1\sigma$,$3\sigma$
and $5\sigma$ regions of $\tilde{\kappa}$ are shown. In obtaining these results we have set $\kappa=1$. 
As it can be seen, the $1\sigma$ region of $\tilde{\kappa}$ is achieved using $\approx 60$ fb$^{-1}$ of the
integrated luminosity of data.

We should mention that our estimation is almost raw due to the lack of 
including the effects of parton shower, hadronization, detector simulation and final state
reconstruction. Including such components with a more detailed analysis 
is beyond the scope of the current analysis and must be performed by 
the experimental collaborations. However, we emphasize that the above
issues may affect the results at a considerable level.

\section{Electric dipole moment analysis}

Beside the high energy studies in the search for the top quark Yukawa couplings,
the low energy probes like electric dipole moments also provide the possibility
to constrain the scalar $(\kappa)$ and pseudoscalar $(\tilde{\kappa})$ components
of Higgs-top couplings. There are already several studies to probe 
$\kappa$ and $\tilde{\kappa}$ using the electron and neutron etc. electric
dipole moments \cite{limits2},\cite{edmhiggs2}.
In \cite{sonit1},\cite{sonit2}, it has been shown that the neutral Higgs boson
exchange with CP-violating couplings at the order of one-loop generates 
sizeable electric and chromo-electric dipole monets for the top quark.
Our main aim in this section is to derive bounds on the $\kappa$ and $\tilde{\kappa}$
parameters using the future achievable bound on the top quark EDM. 
It has been shown that the future electron-positron collider 
would be able to set an upper limit of $10^{-19}$ e.cm directly 
on the EDM of the top quark. In the following, we extract bounds on 
$\kappa$ and $\tilde{\kappa}$ using this upper limit on the top EDM.

\begin{figure}
\centering
\includegraphics[width=12cm,height=10cm]{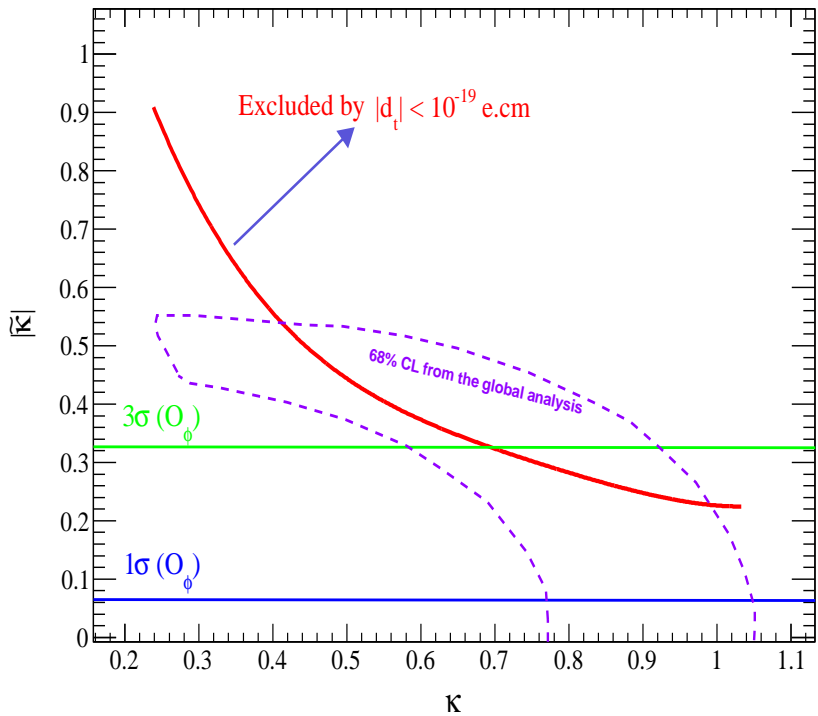}
\caption{$68\%$ CL allowed region from the global analysis taken from \cite{th7} (violet dashed-curve), the $1\sigma$
and $3\sigma$ allowed region obtained using the $O_{\phi}$ observable with 30 fb$^{-1}$ of the LHC data at
14 TeV. The red solid curve shows the allowed region that could be obtained by the future upper limit on the top quark EDM ($d_{t}<10^{-19}$ e.cm).}
\label{global}
\end{figure}

The EDM of a spin-$1/2$ fermion can be defined using the form factor
decomposition of the electromagnetic current $j_{\mu}$:
\begin{eqnarray}
 <p',s'|j_{\mu}(x)|p,s>~\propto~ \bar{u}(p',s')[f_{1}(q^{2})\gamma_{\mu}+f_{2}(q^{2})\sigma_{\mu\nu}\gamma_{5}q^{\nu}+...]u(p,s)
\end{eqnarray}
where $q = p'-p$, $f_{1}(q^{2}=0)$ points out the electric charge of the fermion and $f_{2}(q^{2}=0)$ is the fermion
electric dipole moment. More precisely: $d_{f} = -f_{2}(q^{2}=0)$.
In the non-relativistic limit the EDM of a fermion with spin $\vec{s}$ appears in the interaction of the fermion with an
external electric field $\vec{E}$ according to the Hamiltonian of $H = -d_{f}\vec{E}.\hat{s}$.
Therefore, in the language of Lagrangian, the EDM of a spin $1/2$ particle is defined by the effective Lagrangian
\begin{eqnarray}
\mathcal{L} = -\frac{i}{2}d_{f}\bar{\psi}\sigma_{\mu\nu}\gamma_{5}\psi F^{\mu\nu}
\end{eqnarray}
In general for a theory of fermion $\psi_{f}$ interacting with other heavy fermions $\psi_{i}$
and heavy scalars $\phi_{k}$'s with masses $m_{i}, m_{k}$ and charges $Q_{i}, Q_{k}$ the interaction which
consists of CP violation is given by
\begin{eqnarray}
\mathcal{L} = -\sum_{i,k}\bar{\psi_{f}}[G_{L,ik}P_{L}+G_{R,ik}P_{R}]\psi_{i}\phi_{k}~+~h.c.
\end{eqnarray}
where $P_{L,R}$ are left-handed
and right-handed projection operators. It is notable that $\mathcal{L}$ violates CP invariance 
if $Im(G_{L,ik}G_{R,ik}^{*}) \neq 0$. The one loop EDM of the
fermion $f$ in this case is given by \cite{sonit1},\cite{sonit2},\cite{edmoriginal},\cite{pran}:
\begin{eqnarray}\label{dt}
|d_{t}| = \sum_{i,k}\frac{m_{i}}{ 16\pi^{2}m_{k}^{2}} Im(G_{L,ik}G_{R,ik}^{*})
\times (Q_{i}I_{1}(\frac{m_{i}^{2}}{m_{k}^{2}})+Q_{k}I_{2}(\frac{m_{i}^{2}}{m_{k}^{2}}))
\end{eqnarray}
where because of the charge is conserved at the vertices, $Q_{k} = Q_{f} - Q_{i}$.
In Eq.\ref{dt},
$I_{1,2}(r)$ have the following form:
\begin{eqnarray}
I_{1}(r) = \int_{0}^{1} dx \frac{x^{2}}{1-x+rx^{2}}~,~
I_{2}(r) = \int_{0}^{1} dx \frac{x(1-x)}{1-x+rx^{2}}
\end{eqnarray}
In our case, $\psi_{i}$ and $\psi_{f}$ are the same (top quark) and $\phi$ is the Higgs field so
$Q_{k}=0$.  
As we mentioned  previously the EDM of the top quark can be explored in the
future electron-positron collider in the $e^{-}e^{+}\rightarrow t\bar{t}$ process.
It has been shown that the top quark EDM can be explored down to $10^{-19}-10^{20}$ e.cm.
After replacing the parameters and the limit of $|d_{t}|<10^{-19}$ e.cm the excluded region
of $\kappa$ and $\tilde{\kappa}$ is shown in Fig.\ref{global}. Stronger bound on the 
pseudoscalar component is achieved when we are close to the SM i.e. $\kappa = 1$.
In other word, the top EDM analysis gives stronger exclusion limit for larger values
of pure scalar type Yukawa coupling. At the value of $\kappa=1$, the upper limit of
$10^{-19}$ e.cm leads to exclude any value of $\tilde{\kappa}$ above 0.2. 
Stringent bound of the order of $10^{-21}$ e.cm on top quark EDM excludes 
$\tilde{\kappa}$ above 0.01 at $\kappa=1$. The top quark EDM allows an admixture
of scalar and pseudoscalar Higgs-top coupling. We find that the limit on $\tilde{\kappa}$
is negligibly sensitive to the Higgs boson mass. The change of limit on $\tilde{\kappa}$
 due to shifting the Higgs mass with $\pm 5$ GeV is less than $0.1\%$

For comparison, the allowed region on the plane of  $(\kappa,\tilde{\kappa})$ 
that has been extracted from a global fit on 42 observables such as $H\rightarrow WW^{*}\rightarrow 2l2\nu$,
$H\rightarrow ZZ^{*}\rightarrow 4l$, $H\rightarrow 2\gamma$, etc. The details of the analysis 
and list of all observables can be found in \cite{th7} and references therein.
In addition, we depict the $1\sigma$ and $3\sigma$ regions that can be obtained 
by 30 fb$^{-1}$ of LHC data using the proposed asymmetry-like observable $O_{\phi}$ in the previous section
in Fig.\ref{global}. The $O_{\phi}$ observable is not sensitive to $\kappa$ but is able to 
provide strong limits on $\tilde{\kappa}$. The reachable $1\sigma$ region using this 
observable is $\tilde{\kappa}<0.06$ for almost any value of $\kappa$.

Finally, we point out that the pseudoscalar coupling $\tilde{\kappa}$ contributes to the electron EDM
electric dipole through a two-loop diagram \cite{electronEDM1}. The experimental 
upper bound on the electron EDM provides also stringent limit on the $\tilde{\kappa}$
coupling. The recent upper limit of $8.7\times 10^{-29}$ e.cm \cite{electronEDM2} leads to 
an upper limit of $\tilde{\kappa}<0.01$.
However, as discussed in \cite{th1} direct probes for $\tilde{\kappa}$ are necessary since 
experimentally there is no measurement on the electron Yukuwa coupling and there might be other
contributions  (from BSM such as supersymmetry) to the electron EDM that could lead to a cancellation
of the two-loop top contribution.

\section{Conclusions}

The associated production of the Higgs boson with a pair of 
top quarks at the LHC enables us to explore the Higgs-top Yukawa 
coupling directly. In general, the Higgs-top interaction may consist 
of pseudoscalar component ($\tilde{\kappa}$) in addition to only the scalar part ($\kappa$).
In this paper, based on the angular distributions of the final state
particles in the $pp\rightarrow t\bar{t}H$  process we have suggested a
new asymmetry-like observable $O_{\phi}$ to explore the Higgs-top coupling. 
We show that value of $O_{\phi}$ is sensitive to the pseudoscalar 
component of Higgs-top couplings and receives negligible contribution from 
the scalar coupling. Therefore, it can be a good observable to 
distinguish between the scalar and pseudoscalar terms.
We find that the value of $O_{\phi}$ drops for the events with boosted Higgs boson and
it takes its largest value when there is no cut on the Higgs transverse momentum.
The value of cut on Higgs transverse momentum at which $O_{\phi}=0$ is 
also a sensitive observable to $\tilde{\kappa}$. It is shown that 
applying the common kinematic cuts similar to what used by the CMS experiment 
on the final state particles does not distrub the trend of $O_{\phi}$ versus the 
cut on the Higgs boson $p_{T}$. Applying the kinematic cuts leads to significant increment in 
the amount of $O_{\phi}$. Finally, we find that the $1\sigma$ region of
$\tilde{\kappa}$ is accessible using $O_{\phi}$ with less than 100 fb$^{-1}$
of LHC data at the center-of-mass energy of 14 TeV. 
In the second part of this work we concentrate on the fact that the presence of the 
CP violating component of the Higgs-top coupling ($\tilde{\kappa}$) leads to a sizeable
EDM for the top quark that is achievable by the future $e^{-}e^{+}$ collider. 
We show that a bound of  $10^{-19}$ e.cm on the top quark EDM excludes 
a significant range of $\kappa$ and $\tilde{\kappa}$. In particular, for the
values of $\kappa$ close to the SM ( $\kappa =1$) reasonable bounds 
can be obtained on $\tilde{\kappa}$. Using an upper bound of $10^{-19}$ e.cm 
on the top EDM excludes any value of the pseudoscalar component $\tilde{\kappa}$
above 0.2.

\end{document}